%% file: tmlr.tex
\documentclass[10pt]{article} 
\usepackage[table,dvipsnames]{xcolor}
\usepackage[preprint]{tmlr}

\usepackage{longtable}
\usepackage{fancyhdr}

\input{math_commands.tex}

\usepackage{hyperref}
\usepackage{url}

\title{Extending the OWASP Multi-Agentic System Threat Modeling Guide: Insights from Multi-Agent Security Research}

\author{\name Klaudia Krawiecka \email kkrawiecka@acm.org \\
      \addr Association for Computing Machinery (ACM)\\
      \AND
      \name Christian Schroeder de Witt \email cs@robots.ox.ac.uk\\
      \addr Department of Engineering Science \\
      University of Oxford}

\def\month{MM}  
\def\year{YYYY} 
\def\openreview{\url{https://openreview.net/forum?id=XXXX}} 

\begin{document}

\maketitle

\begin{abstract}
We propose an extension to the OWASP Multi-Agentic System (MAS) Threat Modeling Guide, translating recent anticipatory research in multi-agent security (MASEC) into practical guidance for addressing challenges unique to large language model (LLM)-driven multi-agent architectures. Although OWASP's existing taxonomy covers many attack vectors, our analysis identifies gaps in modeling failures, including, but not limited to: reasoning collapse across planner–executor chains, metric overfitting, unsafe delegation escalation, emergent covert coordination, and heterogeneous multi-agent exploits. We introduce additional threat classes and scenarios grounded in practical MAS deployments, highlighting risks from benign goal drift, cross-agent hallucination propagation, affective prompt framing, and multi-agent backdoors. We also outline evaluation strategies, including robustness testing, coordination assessment, safety enforcement, and emergent behavior monitoring, to ensure complete coverage. This work complements the framework of OWASP by expanding its applicability to increasingly complex, autonomous, and adaptive multi-agent systems, with the goal of improving security posture and resilience in real world deployments.
\end{abstract}

\section{Introduction}

The emerging field of \textit{multi-agent security} (MASEC)~\citep{dewitt2025} is the science and practice of anticipatory security for AI ecosystems: proactively identifying, modelling, and mitigating vulnerabilities before they manifest, guided by what is mathematically or physically possible rather than limited to observed incidents or existing infrastructure. It goes beyond detecting known failure modes to predicting novel risks from emergent behaviours in open-ended agent interactions. Spanning from theoretical modelling to deployable defences, MASEC complements \textit{multi-agent systems security} in its interest in real-world deployments, but distinguishes itself through its anticipatory ambition, its extension to open-ended AI ecosystems beyond existing cyber-physical systems, and its integration of sociotechnical contexts where agents engage with humans and institutions.

This work applies the MASEC framework to extend and operationalize the OWASP Multi-Agentic System (MAS) Threat Modeling Guide~\citep{owasp2024mas}, illustrating how the anticipatory, interaction-focused methodology of MASEC already influences the design of real-world security standards. We identify taxonomic gaps, propose additional threat classes grounded in empirical multi-agent AI research, and outline evaluation strategies aimed at improving the resilience of deployed MAS against both known and emergent adversarial behaviors. The following sections outline the proposed threat categories, illustrate how they manifest in practice, and compare their coverage within the current OWASP taxonomy.

\raggedbottom

\subsection{Terminology: Agent Roles in Multi-Agent Systems}

In this document, we refer to common functional roles observed in multi-agent system architectures (e.g., AutoGen, Reflexion, BabyAGI, and Toolformer). Although these agentic frameworks do not explicitly define the aforementioned roles, we adopt the following terminology to support consistent threat modeling:

\begin{itemize}
  \item \textbf{Planner/Orchestrator (inc. Subplanner)}: An agent responsible for decomposing goals into actionable plans or subtasks. Planners often initiate delegation to other agents.
  \item \textbf{Executor}: An agent that carries out specific actions or tool invocations in service of a plan. It may rely on APIs, external tools, or environments.
  \item \textbf{Verifier}: An agent that \textit{passively} evaluates the validity, safety, or accuracy of the outputs produced by other agents, often acting as a quality control layer.
  \item \textbf{Refiner}: An agent that actively \textit{modifies} the outputs produced by other agents, often acting as active quality assurance (in contrast to the passive Verifier).
\end{itemize}


\section{Proposed Extensions}

Table~\ref{tab:owasp-comparison} outlines the coverage and overlap between existing and new threats presented in the OWASP framework. 
Table~\ref{tab:mas-owasp} outlines additional threat categories and vectors observed in multi-agent systems, which we propose as extensions to the Maestro taxonomy. Each proposed category includes a threat description, a vector, and an example scenario.

\begin{center}

\begin{longtable}[c]{|p{0.15\textwidth}|p{0.18\textwidth}|p{0.25\textwidth}|p{0.35\textwidth}|}

\caption{Coverage of proposed threat categories in the OWASP Multi-Agent Threat Taxonomy} \\
\hline
\label{tab:owasp-comparison}\textbf{Proposed Threat Class} & \textbf{OWASP Coverage} & \textbf{Key Differentiator} & \textbf{Why It Matters in Multi-Agent Systems} \\
\hline 

\textbf{Reasoning Collapse} & \cellcolor{Red!20}Not Covered & The guide does not model breakdowns in stepwise reasoning across planner-executor chains. & Flawed reasoning structure or logic within/among agents. Incoherent plans or logical errors can propagate unchecked, resulting in task failure or tool misuse. \\
\hline

\textbf{Metric Overfitting} & \cellcolor{Red!20} Not Covered & Evaluation-stage nor design-stage metric gaming is not addressed in Maestro's layers. & Agents may prioritize optimizing score signals rather than achieving correct or safe outcomes. \\
\hline

\textbf{Unsafe Delegation Escalation} & \cellcolor{Red!20}Not Covered & Role inheritance or unclear access boundaries are not addressed in delegation threats. & A verifier agent may inadvertently gain executor powers, creating implicit privilege escalation. \\
\hline

\textbf{Evaluation Framework Failures} & \cellcolor{Red!20} Not Covered & The framework does not treat flawed benchmarks or test-time logic as attack surfaces. & Systems may reinforce invalid behaviors if flawed metrics reward hallucinated or biased results. \\
\hline

\textbf{Delegation Pressure Exploits} & \cellcolor{Red!20}Not Covered & Coercive overrides from planning agents are not discussed as a unique failure mode. & A planner may suppress verifier objections and force execution of risky plans. \\
\hline

\textbf{Affective Prompt Framing} & \cellcolor{Red!20}Not Covered & Persuasive or stylistic prompt framing is not modeled as a manipulation vector. & An agent may bias others using tone, confidence, or emotional appeal, bypassing standard checks. \\
\hline

\textbf{Emergent Covert Coordination} & \cellcolor{Red!20}Not Covered & OWASP does not model emergent symbolic protocols or adaptive learning behavior used to evade filters or coordinate covertly. & Multi-agent systems can develop encodings or probing strategies that exploit safety layers over time, even without explicit compromise or instruction injection. \\
\hline

\textbf{Heterogeneous Multi-Agent Exploits} & \cellcolor{Red!20}Not Covered & OWASP does not model threats that arise only from orchestrated coordination across individually safe agents with divergent policies. & Adversaries can split tasks among compliant agents to bypass safeguards; attribution becomes difficult and traditional single-agent monitoring fails. \\
\hline

\textbf{Context distortion from active security monitoring (refinement)} & \cellcolor{Red!20}Not Covered & OWASP does not cover threats arising from active monitoring techniques & Agent communication context distortion due to context loss through sanitisation or active paraphrasing (the latter used to suppress covert coordination)\\\hline

\textbf{Multi-Agent context confusion} & \cellcolor{Red!20}Not Covered & OWASP does not cover threats arising from context confusion when a single agent operates on context generated from multiple agents with different roles & A verifier agent needing to separate contexts from different agent communications on a message board\\\hline

\textbf{Goal Drift in Delegated Chains (Telephone Game)} & \cellcolor{Yellow!20}Partially (Intent Manipulation) & The framework discusses intent manipulation, but not benign drift between agents in the \textit{same} hierarchy (in contrast to Reasoning Collapse). & Multi-agent chains may subtly shift task definitions, leading to unintended actions despite no malicious input (e.g., models following different principles, sharing different context or understanding) \\
\hline

\textbf{Hallucinated Inference} & \cellcolor{Yellow!20}Partially (Tool Misuse, Memory Poisoning) & Hallucinations are treated as isolated events, not multi-agent propagation risks. & Incorrect outputs may cascade through systems if trust and verification protocols are weak. \\
\hline

\textbf{Emergent Collusion} & \cellcolor{Yellow!20}Partially (Cascading Trust Failure) & The guide warns about trust loss but not cooperation for metric exploitation or mutual validation. & Agents may co-adapt outputs to pass evaluations or reduce rejections without external grounding. \\
\hline

\textbf{Trust Misuse Between Legitimate Agents} & \cellcolor{Yellow!20}Partially Covered (Trust Exploitation, Rogue Agents) & OWASP models deception by compromised or spoofed agents but does not address goal-driven misreporting by agents acting within valid roles. & Agents may strategically overstate success or omit uncertainty to maintain trust or optimize local rewards, leading to unsafe but seemingly cooperative behaviors. \\
\hline

\textbf{Multi-Agent backdoors} & \cellcolor{Yellow!20} Partially Covered & OWASP does not cover multi-agent threats arising from model backdoors that specifically compiled to enable multi-agent threats with model copy agents & An agent featuring a secure steganographic capability compiled into its model weights that allows it to conduct covert symbolic coordination with other agents\\\hline

\textbf{Emergent insider threats} & \cellcolor{Yellow!20} Partially Covered & OWASP does not cover multi-agent threats emerging spontaneously from within the system due to misaligned or competitive objectives & Agents become incentivised to use worst-case exploits against other agents due to misaligned or partially competitive objectives, e.g. jailbreaking attacks against verifier agents\\\hline


\textbf{AI supply chain social engineering} & \cellcolor{Yellow!20} Partially Covered & OWASP does not cover threats arising from AI-based social engineering attacks on human factors in AI supply chains  & Mass-scale, long-term manipulation or blackmailing of open source software maintainers as in the xz utils approach using disinformation or AI-generated messages\\\hline

\end{longtable}

\end{center}


\begin{center}
\begin{longtable}[c]{ | p{0.15\textwidth} | p{0.18\textwidth} | p{0.25\textwidth} | p{0.33\textwidth} |}
\caption{Proposed new threat categories with Multi-Agent examples as extensions to OWASP's Multi-Agent Threat Taxonomy} \\
\hline
\textbf{Threat Class} & \textbf{Description} & \textbf{Threat Vector} & \textbf{Example Scenario (Multi-Agent Focus)} \\
\hline

\label{tab:mas-owasp}\textbf{Reasoning Collapse} & Breakdown in chain-of-thought, logic, or planning across agent steps. & Misuse of planner tools, incoherent intermediate outputs. & A planner agent passes vague subgoals to an executor, which attempts invalid tool actions due to unclear/underspecified intent. \\
\hline

\textbf{Metric Overfitting} & Agents overfit to evaluation metrics without solving actual tasks. & Prompt engineering exploits, metric misalignment. This can happen at the systems design level if security metrics insufficiently capture true intents (Goodheart's law). & A chain of agents returns outputs with specific keywords to pass evaluation scripts. \\
\hline

\textbf{Unsafe Delegation Escalation} & Subagents gain unintended elevated privileges due to subgraph permission design flaws. & Role misconfiguration, implicit permission inheritance. & A verifier starts executing tasks due to inherited permissions. \\
\hline

\textbf{Evaluation Framework Failures} & Flawed or outdated evaluation metrics reinforce unsafe behavior. & Ambiguous success criteria, hallucinated correctness. & A verifier accepts hallucinated answers due to vague benchmark definitions. \\
\hline

\textbf{Delegation Pressure Exploits} & Higher-tier agents override downstream agents. & Prompt suppression, system-level override. & A planner forces task execution despite verifier objections. \\
\hline

\textbf{Affective Prompt Framing} & Agents influence decisions through stylistic language, fictional or hypothetical scenarios. & Authority mimicry, emotional tone. & An executor uses confident language that biases the verifier into approving faulty output. \\
\hline

\textbf{Emergent Covert Coordination} & Agents autonomously develop symbolic protocols or adapt behaviors to covertly coordinate, bypass safety filters, or encode intent. These behaviors emerge without explicit programming. & Multi-round prompt shaping, token-based flagging, steganography, learned filter probing, covert signaling. & A planner and verifier evolve a shared flag convention in outputs to bypass a content filter that only evaluates explicit prompt phrasing. \\
\hline

\textbf{Heterogeneous Multi-Agent Exploits} & Attackers coordinate multiple agents with different capabilities, each compliant in isolation, to bypass safety mechanisms collectively. & Delegated task splitting, policy evasion via complementary agents, cross-model chaining. & A frontier LLM with refusal policies handles code synthesis, while a jailbroken model crafts unsafe phrasing; together they produce vulnerable outputs undetected by either model alone. \\
\hline

\textbf{Context Distortion from Active Security Monitoring (refinement)} & Agent communication context distortion due to context loss through in/output or communications refinement.  & Active security monitoring of in/outputs or communications, including sanitisation and filtering, or paraphrasing (the latter used to suppress covert coordination). & A dynamic firewall sanitises a natural language query in a way that drops subtle query context, leading to wrong, inaccurate, or sensitive data being returned from the database. \\\hline

\textbf{Multi-Agent Context Confusion} & Reasoning failures over multi-agent contexts & Innate foundation model weaknesses in handling context levels. & A verifier agent needing to separate contexts from different agent communications on a message board. \\\hline

\textbf{Goal Drift in Delegated Chains (Telephone Game)} & Planner's intent shifts through agent-to-agent delegation. & Poor prompt transmission (inc. prompt interfacing, formatting) . & A subplanner alters strategy, and the executor completes an action misaligned with the original goal. \\
\hline

\textbf{Hallucinated Inference} & Agents generate confidently incorrect outputs that propagate (Could trigger Reasoning Collapse and Goal Drift) & Lack of grounding, recursive generation flaws. & An executor hallucinates a function; the verifier accepts it based on prior trust. \\
\hline

\textbf{Emergent Collusion} & Agents reinforce each other’s outputs for mutual gain. & Reciprocal approvals, echo validation. & A verifier always accepts the output of an executor to maintain high success rates of tasks. This includes the planner-executor-verifier coordination to maximize their combined reword or global task score. \\
\hline

\textbf{Trust Misuse Between Legitimate Agents} & A non-compromised agent misleads another agent within the system to maximize local goals or optimize rewards, without external compromise or identity spoofing. & Overconfident outputs, fabricated success claims, omitted caveats. & An executor agent overstates task success to the verifier to avoid rollback, preserving its local performance metric despite introducing downstream risk. \\
\hline

\textbf{Multi-Agent Backdoors} & Backdoor circuits hidden in agent model weights that enable agents to orchestrate multi-agent threats & Agent model backdoors either learnt from poisoned data, or compiled into the model weights & An agent based on open-source model weights featuring a hidden secure steganographic capability compiled into its model weights that allows it to conduct covert symbolic coordination with other agents based on the same model \\\hline

\textbf{Emergent insider threats}  & Agents become incentivised to use worst-case exploits against other agents due to emergence under misaligned or partially competitive objectives  & Optimisation under misaligned optimisation objectives or unexpected environment shifts (including in-context learning) & Agents perform jailbreaking attacks against verifier agents due to subtle misalignments in multi-objective optimisation objectives. \\\hline


\textbf{AI supply chain social engineering} & Threats arising from AI-based social engineering attacks on human factors on AI supply chains  & AI-generated disinformation or communications spread by bots or humans & Mass-scale, long-term manipulation or blackmailing of open source software maintainers as in the xz utils approach using disinformation or AI-generated messages\\
\hline

\end{longtable}

\end{center}

\section{Testing Framework(s)}

\subsection{Robustness}
An ongoing work from \cite{owotogbe2025} suggests 'chaos engineering' to stress-test LLM-based multi-agent systems. This means deliberately injecting failures, for example, by introducing communication delays, or corrupting messages. This framework could be used to simulate agent failures and communication breakdowns in LLM multi-agent setups. This helps ensure the whole system remains reliable even if some agents behave unexpectedly.

Another idea is to test agentic communication topology against malicious interference. The NetSafe framework \cite{yu2024netsafe} could  evaluate safety under targeted attacks in a network of LLM-driven agents. It suggests injecting malicious content (misinformation, biased, or harmful prompts) into certain 'attacker' agents and measures how the bad information spreads through various network structures.

Beyond specific frameworks, we could look at simulations where agents might be placed in certain scenarios with missing information, random interruptions, or noisy inputs to see if they still achieve desired goals. The purpose is to ensure the multi-agent system can degrade gracefully, which means that if one agent fails or provides a wrong output, others should detect and correct it (akin to fault tolerance).

\subsection{Coordination Evaluation}

A central promise of multi-agent LLM systems is improved performance through coordination – multiple agents working together should accomplish tasks more effectively than any could alone. Evaluating coordination involves measuring how well agents communicate, synchronize, and complement each other’s actions.

The most direct metric is success on cooperative tasks. Benchmarks from multi-agent reinforcement learning and board games are used to test LLM-based agents. For example, the Star-Craft Multi-Agent Challenge~\cite{samvelyanStarCraftMultiAgentChallenge2019g} (a cooperative card game requiring communication under partial information) has been a standard for coordination in AI (though typically with RL agents). More recently, VendingBench~\cite{lesswrong} is a virtual environment where two or more agents with a shared goal must operate a vending machine together. The evaluation checks if they can coordinate actions to acquire resources (coins, items) without mishaps. Such scenarios yield metrics like task completion rate (did the team achieve the goal?), efficiency (time or steps taken), and resource utilization (did they waste actions due to poor coordination?).

In addition, agents could be assessed on how 'in sync' they are. In cooperative settings, one can measure the agreement or consistency among agents’ decisions. For instance, in a hidden-role game environment, \citet{curvo2025traitors} define a Faithful Agreement Score ('quantifies consensus among faithful agents') and Traitor Agreement Score ('measures how consistently traitors vote as a single unified group') to see how consistently each group votes together. High agreement within teams indicates effective coordination or collusion, whereas divergence might signal miscommunication. Similarly, they could measure the communication overhead required through this game. This study has tested how different models handle deceptions between each other.

\subsection{Safety}

TrustAgent \cite{hua-etal-2024-trustagent} evaluates safety via a three-stage process: before an agent makes a plan, it 'prepends' safety knowledge to its context, during the plan, it uses special prompting to steer away from unsafe choices, after the plan, it performs checks and self-edits. In evaluations across multiple domains, this framework successfully identified and mitigated potential dangerous actions the agent was about to take, thereby reducing the occurrence of unsafe outputs. For instance, if an agent’s plan involved accessing private user data, the 'constitution layer' would flag and alter that. The experiments showed not only improved safety compliance, but interestingly, also a boost in the helpfulness of the agent. This demonstrates that structured safety enforcement can be evaluated by comparing agent behavior with and without the safety strategies, measuring metrics like safety violations prevented and task success retained.

As mentioned earlier, NetSafe \cite{yu2024netsafe} examines how a network of agents can resist unsafe content propagation. One safety dimension this study measured was hallucinations and aggregation safety, referring to phenomena where one agent’s hallucinated misinformation gets accepted and amplified by others. NetSafe introduced static metrics (graph-based measures of network resilience) and dynamic metrics (like the drop in task performance when an attack is introduced). These metrics were validated by showing strong alignment with actual outcomes in the experiments, e.g., a network that scored better on the static safety metric indeed suffered less performance degradation when malicious prompts were injected. This kind of evaluation is practical for organizations deploying many agents since it helps decide how to connect agents safely (e.g., limit which agents can talk to which others) by quantifying the risk of a 'contagion' of errors or harmful content.

Overall, multi-agent setups can include redundancy for safety. For example, one agent can be assigned as a 'verifier' to critique or veto another agent’s potentially harmful decision. An evaluation methodology here is to use challenge scenarios, e.g., have one agent suggest an unsafe action and see if the second agent catches it.

\subsection{Emergent Behaviors}

One of the most interesting aspects of multi-agent systems is the potential for emergent behavior – complex dynamics or capabilities that are not programmed in any single agent, but arise from agentic interactions.

Particioners can (and probably should) create sandbox multi-agent ecosystems or long-term simulations and simply let multiple agents interact, logging everything. Through a simulations like this, we can catch emergent behaviors and we can measure them, for example, novelty metrics (did the system generate new kinds of activity?), complexity metrics (e.g. average length of interaction chains, number of agents involved in a single event), and through human evaluation of plausibility.

In scarce evaluations so far, the emergent behaviors observed (tool use, social planning, secret codes, etc.) have often been impressive but also cautionary. They show that multi-agent systems can evolve beyond their initial design. Therefore, a forward-looking evaluation strategy could be to include long-run simulations in test suites – essentially, 'let’s watch the agents interact for 100 steps and see if anything odd or interesting happens' and have analysts or monitors ready to capture that. As multi-agent AI deployments become persistent (e.g. a fleet of service robots or a network of dialogue agents that continually talk), this kind of ongoing monitoring becomes part of the evaluation pipeline, ensuring that emergent behaviors are caught early and aligned with human intent.


\bibliography{tmlr}
\bibliographystyle{tmlr}


\end{document}

%% file: math_commands.tex
\usepackage{amsmath,amsfonts,bm}

\def\eqref#1{equation~\ref{#1}}

\def\1{\bm{1}}

\DeclareMathAlphabet{\mathsfit}{\encodingdefault}{\sfdefault}{m}{sl}
\SetMathAlphabet{\mathsfit}{bold}{\encodingdefault}{\sfdefault}{bx}{n}

